\documentclass[10pt]{article}
\usepackage{graphicx}
\usepackage{bm}
\usepackage{setspace}

\newcommand\nh {n_{\mathrm H}}
\newcommand\nf {n_{\mathrm F}} 
\newcommand\nsv {n_{\mathrm SV}}
\newcommand\nre {n_{\mathrm RE}}

\newcommand\Nh	 {N_{\mathrm H}}
\newcommand\Nf	 {N_{\mathrm F}} 
\newcommand\Nsv {N_{\mathrm SV}}
\newcommand\Nre {N_{\mathrm RE}}

\begin{document}
\pagestyle{empty}

\title{Field evidence of social influence in the expression of political preferences: 
the case of secessionist flags in Barcelona
}
\maketitle

\begin{center}
\author{Antonio Parravano$^{1,2}$, Jos\'e A. Noguera$^3$, Jordi Tena$^3$ and Paula Hermida$^3$}\\
\vspace{0.3cm}
{\small
{$^1$ Universidad de Los Andes, Centro de F\'{\i}sica Fundamental, 
M\'erida, Venezuela\\}
{$^2$ Instituto de Estudios Sociales Avanzados (IESA-CSIC), C\'ordoba 14004, Spain\\}
{$^3$ Universitat Aut\`onoma de Barcelona, Group of Analytical Sociology and Institutional Design (GSADI), Barcelona, Spain\\}
}
\end{center}

\begin{abstract}
Different models of social influence have explored the dynamics of social contagion, 
imitation, and diffusion of different types of traits, opinions, and conducts. However, 
few behavioral data indicating social influence dynamics have been obtained from direct observation in 
`natural' social contexts. The present research provides that kind of evidence in 
the case of the public expression of political preferences in the city of Barcelona, where 
thousands of citizens supporting the secession of Catalonia from Spain have placed a Catalan flag in their 
balconies and windows. Here we present two different studies. 1) During July 2013 we registered 
the number of flags in 26\% of the electoral districts in the city of Barcelona. We find that there is a 
large dispersion in the density of flags in districts with similar density of pro-independence 
voters. However, by comparing the moving average to the global mean we find that the density of 
flags tends to be fostered in those electoral district where there is a clear majority of 
pro-independence vote, while it is inhibited in the opposite cases. We also show that the 
distribution of flags in the observed districts deviates significantly from that of an equivalent random 
distribution. 
2)  During 17 days around Catalonia's 2013 National Holiday we observed the position at balcony resolution 
of the flags displayed in the facades of a sub-sample of 82 blocks. We compare the `clustering 
index' of flags on the facades observed each day to thousands of equivalent random 
distributions. Again we provide evidence that successive hangings of flags are not independent events but 
that a local influence mechanism is favoring their clustering. We also find that except for the National 
Holiday day the density of  flags tends to be fostered in those facades where there is a clear 
majority of pro-independence vote.
 \end{abstract}

Keywords: social influence, social dynamics, political preferences, social behavior.

\section{Introduction:}

People's decisions, opinions and behavior partially depend on what others decide, think and 
do \cite{Ball12,CialdiniEA04,CialdiniEA98}.
The concept of social influence refers to the fact that in many social contexts the probability of 
an individual acting in a 
given way depends on how many individuals have already acted in that way \cite{HedstromEA10,WattsEA09,SalganikEA09}. 
Therefore the social diffusion of a given 
behavior may be typically affected by the perception people has of how the members of the relevant group are 
behaving. As a mechanism to explain social diffusion, social influence (adjusting to perceived 
collective behavior) has been distinguished from social contagion (start doing A when you contact 
someone doing A), from rational imitation (under uncertainty, do as everybody else does) and 
from social learning (do A when you see A works fine for others) \cite{NogueraEA14,Young09}.
Different models have explored the dynamics of social influence, imitation, and contagion of many 
types of social and political behavior, opinions and traits 
\cite{Aguiar-Parravano-13, BakshyEA12,BalboEA14,BaldiEA14,Ball12,BischoffEA13,ChristakisEA12,CialdiniEA04, DurlaufEA10, 
FlacheEA11,LiuEA10,FalkEA02,LopezEA08,Manski04,NogueraEA14,Young09,WattsEA09,XiaEA13}.
The spread of obesity, 
smoking, alcohol consumption, happiness, divorce, suicide, sexual practices, tastes in music, books 
and movies, altruism, political mobilization, electoral preferences, and many other conducts, 
beliefs and preferences has been modeled in the literature on social influence.

Leaving aside purely theoretical models, most empirical studies to date rely on four different sources 
of evidence: surveys or longitudinal data \cite{AbergEA11,ChristakisEA12,Manzo13,AbrutynEA14},
virtual networks and internet behavior \cite{BakshyEA12,BondEA12,BorgeEA13,Centola10,WattsEA98,SalganikEA09},
historical records \cite{GonzalezEA13,Hedstrom94},
and experiments \cite{CialdiniEA04,MasEA13,DeutschEA55,DiekmannEA11,FortinEA07,MousaidEA13}.
In contrast to the vast majority of studies in this field, in this article 
we present behavioral evidence obtained from direct observation in a particular social context: the 
public expression of political preferences in the city of Barcelona during July and September 2013, 
through the placing of Catalan secessionist flags in balconies and windows.

This paper makes several contributions: first, it provides a case-study of a social influence process 
by direct observation of an objective behavior in a real context; this allows to discard possible 
biases introduced by subjectivity in self-reporting (as in survey evidence), artificial situations 
such as experimental treatments or survey interviews, and virtual behavior in the net as opposed to 
interaction in physical social contexts. Second, our study provides evidence that not only political 
preferences are affected by social influence \cite{BischoffEA13,BondEA12},
but also its public expression; this is important since the literature has widely discussed cases of 
spiral of silence or pluralistic ignorance regarding political preferences \cite{Bicchieri06,CentolaEA05,WillerEA09,KuranEA95},
but large empirical data on such cases are still scarce. 
Third, our research also gives empirical support to the thesis that social influence has a 
strong spatial dimension \cite{ChristakisEA12,Latane81,LiuEA10},
that is, that the probability of an agent being socially influenced by others is higher the smaller 
is the physical distance between them.

The observed behavior (hanging a flag in private households' windows and balconies) is a promising 
candidate to be affected by social influence: it is easily observable for any individual in the area, 
it has a simple binary structure (to hang or not to hang a flag), and a very clear meaning (to express 
support for the Catalan secessionist process). It is a case of influence that results from the 
need to identify oneself as a member of a group in terms of political preferences, and to signal it 
to the world \cite{CialdiniEA04,DeutschEA55}.

However, some clarifications are due. First, we assume that social influence generated by flag-hanging 
would not be affecting political preferences as such, but their expression through the 
successive placing of flags; political support for Catalan independence is most likely to be produced 
by other factors than seeing flags. What we want to determine is whether an individual already 
having pro-independence preferences will tend more or less to hang a flag depending on how many others 
do so. Second, our focus is whether the observed distribution of flags indicates the existence of 
social influence, not which specific social influence mechanism is in place; we show that social 
influence is necessary to explain the observed distribution, but several different mechanisms (or 
combination of mechanisms) might be consistent with the pattern (this is a usual problem of studies 
on social influence: \cite{LopezEA08,NogueraEA14},
however some constraints on the spacial scale of the influence mechanisms can be established. 

Since a preliminary observation of the main streets and avenues of the city in January 2013 suggested that flags tend  
to appear together in clusters, and that there is not a linear relationship between the frequency 
of flags and variables such as voting behavior or income level, we gather systematic observational data 
in order to test whether flags are clustered in a non-random way, and whether its distribution 
is completely explained by voting behavior or a social influence process may also be in place. 
Our hypothesis is that the probability of placing a Catalan flag in a given household's balcony or window 
is correlated with voting for secessionist parties, but also significantly affected by the number 
of neighbors who hang a flag. We test this hypothesis by analyzing two set of data: one at a coarse resolution 
where the unit of observation is one electoral district, and other at a fine resolution 
where the unit of observation is one household's balcony. 
The quantitative characterizations of the state of the system at both the mesoscale and at the microscale, 
as well as the reaction of the system to a the Diada effect are valuable constraints for theoretical models.

\section{The case: flag-hanging behavior and the Catalan secessionist process}

On Catalonia's national day (11 September, the `Diada'), lots of people traditionally hang 
a Catalan flag in their balconies or windows, and remove it the day after. But since the 2012 Diada, 
thousands of flags stay hanged, and others appear. The reason is that many citizens want to 
express their political support for Catalonia's secessionist process started that year by 
the nationalist Catalan government.

From 2005 onwards, the constitutional status of Catalonia within the Spanish state has been subject 
to strong political struggle and discussion. In 2006, the Spanish Parliament significantly cut 
Catalan aspirations in the proposal for a new Autonomy Statutory Act, and in 2010 the Spanish 
Constitutional Court abolished important parts of the Catalan Statutory Act which was approved in 
referendum in 2006. During all this period, political struggle between Catalan and Spanish 
governments on funding and linguistic rights has dominated the political agenda. This situation raised 
massive demonstrations in Barcelona on a yearly basis from 2010, until the Catalan president, 
echoing a widespread social and political mobilization, proposed to celebrate a referendum on 
independence from Spain and called for early Catalan elections in 2012. In this election, 
secessionist parties won a big majority of the Parliament. In 2013 four pro-referendum parties agreed on 
celebrating the referendum on November 9th 2014. The Spanish government has refused to negotiate 
and has announced that the Constitutional Court will ban any referendum. 

In this context and since then, Catalan flags have proliferated in Barcelona's balconies and windows. 
The act of hanging a Catalan flag in a flat's balcony or window has become a very usual way of 
publicly expressing support for `the process' (a term secessionists used to name the political 
and social road to independence). There is little doubt that someone placing a Catalan flag in his 
household's forefront is clearly expressing support for celebrating a referendum, and most likely 
also for independence. The difference is relevant in some cases because the agreed question for the 
proposed referendum includes a third option that Catalonia becomes a non-independent state. 
Besides, two of the pro-referendum parties (Converg\`encia i Uni\'o, CiU and Iniciativa per Catalunya-Verds, 
ICV) are not unanimous on their support to independence. However, all polls show that only a 
very slight fraction of voters (including voters of those two parties) would opt for the `third way' 
rather than for secession or the status quo. Even if not absolutely all the people who hang a flag
might vote for independence, it is quite granted that none of them would vote for the status quo and 
that they all support the referendum. In consequence, and for simplicity, we will use the term 
`pro-referendum' instead of `pro-independence' from now on.

\section{Data and methods}

\subsection{First study}
\begin{figure*}
\hbox{\includegraphics[scale=0.25,angle=0]{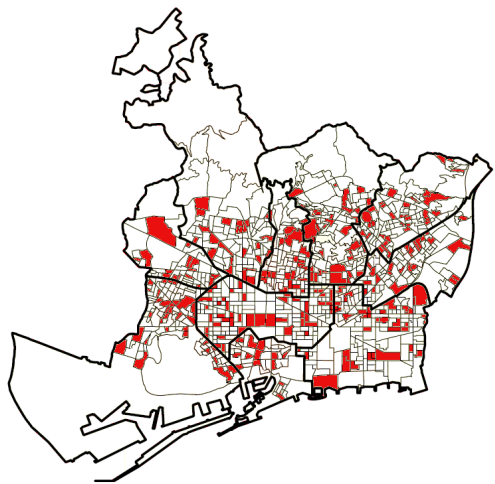}}
\caption{\label{fig:eds}
{\small
{\bf Sample of 276 electoral districts observed in the city of Barcelona} (in red). Thick 
lines indicate administrative district borders; thin lines indicate electoral district borders.}}
\end{figure*}

In order to test whether a social influence mechanism is operating in placing flags, we first observed 
during July 2013 the complete distribution of flags in a representative sample of 276 electoral 
districts (EDs henceforth) in the city of Barcelona (26\% of the total); the sample includes 213,667 
households in which 293,144 voters are registered. EDs are relatively small areas with a mean of 
1,062 registered voters each. The sample was stratified at the quarter level in order to ensure that 
all 76 quarters in the city were adequately represented.
EDs were randomly selected within each quarter and the sample error is 5\%, with p=q=50\% (which 
maximizes sample size). Six EDs were removed 
from the sample because they are industrial or rural areas where very few households are registered and 
the rest of potentially eligible EDs in their quarters were similar. Other two similar cases were 
replaced randomly by other EDs in the same quarters. Figure (\ref{fig:eds}) shows the extension of the selected 
sample in Barcelona's map (see also Table 1).

Only Catalan national flags were registered, in their two usual versions: senyera 
and estelada (with a star). Only one flag per household was counted. Flags in commercial 
or office establishments, as well as in any other non-private households, were not counted.
We analyzed how the density of flags correlates with the level of pro-independence vote in 
the 2012 election for the Catalan Parliament, in order to see if flag-hanging behavior 
tends to be inhibited in those EDs where there is not a clear majority of pro-referendum vote, 
while it is fostered in those where such a majority clearly exists. We also analyzed if the distribution 
of flag densities in the 276 electoral districts appreciably deviates from the expected distribution 
if pro-referendum voters display a flag with a uniform probability. Therefore, at the ED scale we 
characterize i) the non-linear correlation between the flag density and the density of secessionist 
voters; ii) the expected dispersion of flag densities for EDs with similar density of secessionist 
voters; and iii) the departure of the distribution of flag densities from that of an equivalent 
random distribution.

\subsection{Second study}
We selected a sub-sample of sixteen spatial areas in different electoral districts. 
A 2x2 typology was designed and four spatial areas of each type were selected according to two 
criteria: density of flags (under and over the average) and type of street (wide/narrow). Most areas 
consist on the facades at both sides of three consecutive blocks. The 82 block facades differ from 
each other in shape and size. The smallest number of households in a block is 4 and the larger one is 351; 
the average number of households per block is about 59. The data consist on a detailed daily 
record of the position of the flags at household resolution in the facades of the 82 blocks from 4 to 18 
September 2013, followed by two additional observations, one on November 19 and the other on 
December 20; this makes a total of 17 different observations. A total of 4,817 households were observed 
in each of the 17 days.
Note that September 11th is Catalonia's national day (the `Diada'), and it is usual to hang a Catalan 
flag in the balcony that day. The method and criteria for counting flags was the same as in the 
first study. This time the records were taken on templates of the facades of the blocks, where the 
separation between households had been previously identified and drawn. 
This observation provides empirical evidence on the detailed dynamics by which flags `appear' before 
September 11th and `disappear' afterwards, departing from a given pre-existing level of flags. The 
intention was to measure how a global strong stimulus produce different effects on blocks subjected 
to different initial condition and/or different environments. Again, we analyze if the specific 
spatial distribution systematically differs from the level of clustering that would be expectable by 
chance; we then test whether the successive hangings of flags are independent events or there 
is some influence mechanism that favors their clustering. So, at the balcony scale, we characterize 
i) the evolution of the total number of flags in the 82 observed blocks; ii) the distribution of a 
the clustering index (Eq. \ref{eq:Cobs}) in the 82 observed blocks and its departure from that of an equivalent 
random distribution.

\section{Evidence of social influence at the mesoscale (ED level)}{\label{sec:meso}}
During the second and third weeks of July 2013 we recorded once the number of pro-independence flags in all 
EDs of the sample. During this short period of time there were not any special political 
events able to produce noticeable changes in the number of flags, so we assume that the observations were simultaneous. 
These data allows to analyzing the effect of social influence on the density of flags at meso-level resolution. 
Detailed electoral results of the 2012 election for the Catalan Parliament (see Table 1) 
as well as data on average annual income are available for each ED. We assume that pro-referendum 
voters are those that in the 2012 election voted for one of the five referendum parties 
(Converg\`encia i Uni\'o, CiU; Esquerra Republicana de Catalunya, ERC;  Iniciativa per Catalunya 
Verds, ICV; Candidatura d'Unitat Popular, CUP; and Solidaritat Catalana per la Independ\`{e}ncia, SI).
\begin{table}
\begin{center}
\begin{tabular}{c}
\hspace{0.0cm}{\bf Table 1}\\ Aggregated sample data
\end{tabular}\\
\begin{tabular}{cccc}
\hline
\hline
  \vspace{-0.2cm}  &\vline &total & mean per ED\\
   &\vline  &--------&-----------------\\
households$^a$& \vline&213667&  774.2\\
flags& \vline&  5479&   19.9\\
\hline
&&&\\
\vspace{-0.2cm}
2012 election$^a$& \vline&total & mean per ED\\
-----------------&\vline  &--------&-----------------\\
registered electors& \vline&293144& 1062.1\\
pro-referendum vote& \vline&123740&  448.3\\
CIU& \vline& 60504&  219.2 \\
ERC& \vline& 27775&  100.6 \\
ICV& \vline& 24846&   90.0 \\
CUP& \vline&  8183&   29.6 \\
SI & \vline&  2432&    8.8 \\
Abstention & \vline& 85383&  309.4 \\
\hline
 \hline
\end{tabular}
\end{center}
\hspace{2cm} $^{a}${\small Data source: Barcelona’s City Hall statistical department.}
\end{table}

\begin{figure}[ht]
\hbox{\includegraphics[scale=0.7,angle=0]{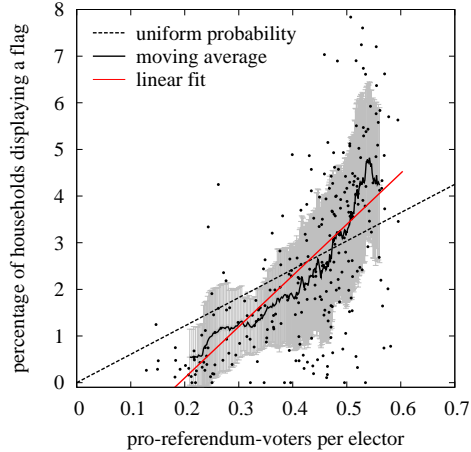}}
\caption{\label{fig:BoV-VS-SoE}{\small
{\bf Percentage of observed households with a flag ($100 \nf/\nh$) as function of the number of pro-referendum 
voters per elector ($\nsv/\nre$)}. The dots give the position of the 276 electoral districts. 
The dotted strike line  $Y_u(X) =6.075 X$ represents the expected value if all pro-referendum voters have 
an uniform probability to hang a flag independently of density of flags in their electoral 
district. The irregular curve corresponds to the moving average of subsets of 30 consecutive data points 
with increasing $\nsv/\nre$ values and the gray error bars show the corresponding  standard 
deviations of the data in each of these subsets . The red strike line is the linear fit to the 276 data points.
}}
\end{figure}

In the 276 observed EDs, 2.6\% of households were displaying a flag when the fieldwork was 
done, but it must be noticed that not all households have balconies or windows at the external facade of the 
buildings, so if only the latter were taken into account the percentage would be higher. 
Let $\nf(i)$, $\nh(i)$, $\nsv(i)$, and $\nre(i)$ be respectively the number of flags, households, 
pro-referendum votes, and registered electors in the electoral district $i$. The dots in Figure 
(\ref{fig:BoV-VS-SoE}) show the relation  $\nf(i)/\nh(i)$ vs $\nsv(i)/\nre(i)$ for the $1 \leq i \leq 276$ EDs. 
That is, the $Y$-axis represents the percentage of householders displaying a flags 
whereas the $X$-axis represents the fraction of registered electors that in 2012 voted for one of the 
pro-referendum parties.  It is evident that there is a large variation of the density of flags in 
electoral districts with similar percentages of pro-referendum voters, although as expected, in average, 
there is a positive correlation between $X$ and $Y$ (the Pearson's correlation coefficient is $r=0.64$). 
For the overall set of EDs, $\Nf/\Nh=0,026$ and $\Nsv/\Nre=0,422$ (see Table 1) where $N_x=\sum_{i=1}^{276} n_x$. 
Therefore, the dashed strike line  $Y_u  = 6.075 X$ represents the expected percentage $Y_u$
of householders displaying a flag  when the fraction of pro-referendum voters is $X$, assuming 
that all pro-referendum householders have an uniform probability $p_u=0.0607$5 to hang a flag 
(dash line in Figure \ref{fig:BoV-VS-SoE}), that is, as if there were no social influence into play. 
It is evident that there are less points above the strike line $Y_u(X)$  than below, and that the 
absence of points is concentrated on the left side of the plot. If the flags in the EDs were randomly 
distributed one would expect that the data were evenly distributed above and below the line 
$Y_u(X)$. Assuming that only pro-referendum voters hang flags with a probability $\Nf/\Nsv=0.0443$, one would 
expect that in the plane $\nf - \nsv$ about half of the EDs fell above the line $\Nf=0.0443 \Nsv$, but only 
118 of the 276 EDs do. Similarly, one would expect that in the plane $\nf/\nh - \nsv/\nre$ shown in 
Figure (\ref{fig:BoV-VS-SoE}) about half of the EDs fell above the line 100 $\nf/\nh =6.075 
\nsv/\nre$, but only 111 of the 276 EDs do. This asymmetry is indicative that a local social 
influence mechanism is intervening in the decision of displaying a flag. The irregular growing curve in 
Figure (\ref{fig:BoV-VS-SoE}) corresponds to the moving average $Y_{m}(X)$ of subsets of 30 
consecutive data points with increasing $\nsv/\nre$ values. This curve helps to visualize the non-linearity 
induced by social influence. At low values of $X$ the EDs tend to have flag densities below 
the expected value $Y_u(X)$ whereas the contrary occurs at high values of $X$. An explanation for this 
phenomenon would be that pro-referendum voters are more likely to place a flag when they perceive 
they are a majority in the neighborhood. In this case, stimulation in the public expression of 
secessionist preferences would be at work in those EDs where there is a clear majority of pro-referendum 
voters, but inhibition would be the case in those EDs where there is not.  Since the social 
influence under consideration is visually mediated, a more precise reasoning would be that pro-referendum 
voters increase their tendency to hang a flag in their balcony when they observe that in their 
environment there is a high density of flags, which usually, but not necessarily, occurs in EDs 
with high proportion of pro-referendum voters. 

In order to characterize the pattern shown in Figure (\ref{fig:BoV-VS-SoE}) here we define two quantities: 
the mean relative dispersion $<\tilde{\sigma}>$ and the social influence index $I_{SI}$. 1) 
The mean relative dispersion $<\tilde{\sigma}>$=0.58 is measured as the mean value of the 
ratio $\sigma(X)/Y_m(X)$, where $\sigma(X)$ is the standard deviation of the subset of the 30 data located 
around $X$ and $Y_m(X)$ is the value of the moving average at $X$. 2) The social influence index is 
measured as $I_{SI}=p_{lf}/(100 p_{u})-1=0.81$, where $p_{lf}=10.1$ is the slope of the linear fit 
to the data (red line in Figure (\ref{fig:BoV-VS-SoE})) and 100 $p_{u}=6.075$ is the slope 
$dY_u/dX$ of the uniform probability line (black dashed line in Figure (\ref{fig:BoV-VS-SoE})). 
These quantities are useful to characterize the observed pattern and to check the results from 
simulation models. For example, If  $<\tilde{\sigma}>$ and $I_{SI}$ are both small, the data points are 
all close to the uniform probability strike line $Y_u(X)$, the correlation is $r \simeq 1$, and social 
influence has a negligible effect on the observable variable. In our case $I_{SI}>0$ which is 
most likely due to the fact that a mechanism of social influence is encouraging the adoption of the 
behavior under observation Y (in our case the density of flags in the ED) when the fraction of 
potential adopters $X$ increases  (in our case the fraction of pro-referendum voters in the ED). 
If instead  $I_{SI}<0$ , it is expected that a mechanism of social influence is discouraging the 
adoption of the behavior $Y$ as $X$ increases. It is to be noticed that similar trends occurs when 
the data are plotted in the plane $\nf - \nsv$, but the positive curvature of the moving average is 
less pronounced, suggesting that what is perceived as social pressure is more the density than 
the absolute number of flags in the ED.

\begin{figure}
\hbox{\includegraphics[scale=0.7,angle=0]{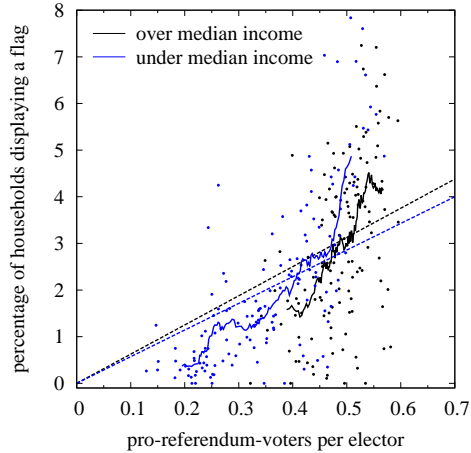}}
\caption{\label{fig:BoV-VS-SoE_2rentas}{\small
{\bf Same as Figure (\ref{fig:BoV-VS-SoE}) but for two equal size subsets of annual average household income levels}. 
Values are attributed to each ED from income level estimations at quarter level. The 
average annual income in the sample is 19,750 Eu and the median is 0.89 of this average value. 
Blue, and black symbols correspond to EDs with income under and over the median value, 
respectively. The moving average uses subsets of 20 data points instead of 30 as in Figure (\ref{fig:BoV-VS-SoE}).
}}
\end{figure}
\begin{figure}
\hbox{\includegraphics[scale=0.65,angle=0]{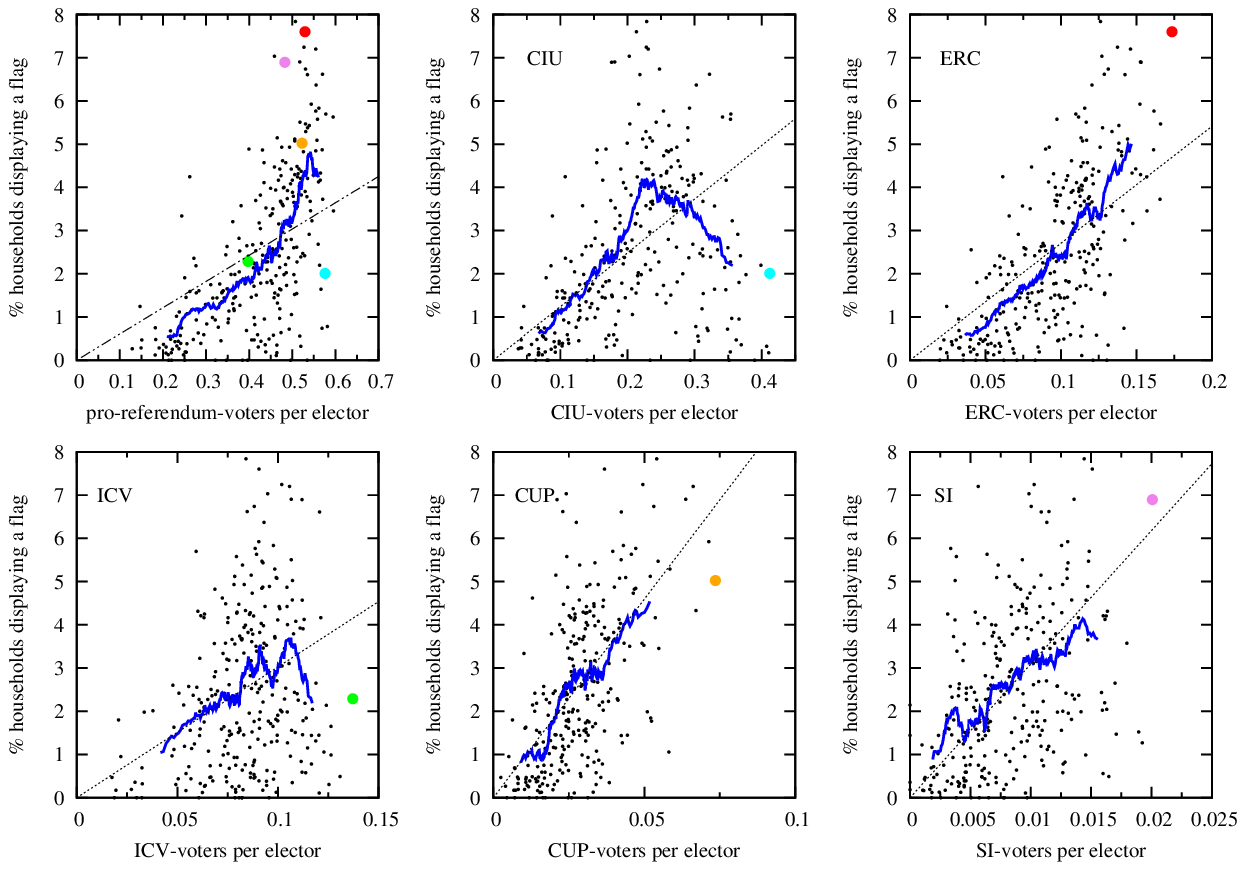}}
\caption{\label{fig:cincopartidos}{\small
The top left plot is the same as Figure (\ref{fig:BoV-VS-SoE}) but indicating with coloured points the 
EDs in which the pro-referendum parties obtained the maximum number of votes; cyan for 
CiU; red for ERC; green for ICV; orange for CUP, and violet for SI. The remaining five plots show 
the correlation between flag density and the fraction of registered electors who voted in the 2012 
election for the Catalan Parliament for the political party indicated in the label. 
The moving average uses subsets of 30 data points.
}}
\end{figure}

Figure (\ref{fig:BoV-VS-SoE_2rentas}) shows the same relation as Figure (\ref{fig:BoV-VS-SoE}) 
but for two subsets of annual average household income taken from the Barcelona's City Hall statistical 
Department. The firs subset contain the half of the EDs having incomes below the median 
value (blue dots and curves characterized by quantities $<\sigma>$=0.62, $I_{SI}=1.15$ and r=0.74). The second 
subset contain the EDs with incomes over the median (black dots and curves characterized 
by $<\sigma>$=0.50, $I_{SI}=1.60$ and $r=0.51$). It is interesting to note that the effect of local social 
influence manifests even more clearly in these subsets, particularly in the high income subset. 
This result rules out the possibility that the positive curvature of the moving average curve in Figure 
(\ref{fig:BoV-VS-SoE}) was due to dissimilar behavioral dispositions in different income segments. 
It is interesting to note that the EDs in the over-median subset have a much narrower and higher $X$ 
range than the EDs in the under-median set, however it is out of the scope of this study to analyze the 
reasons and consequences of this asymmetry. Nevertheless, it is important to point out that this 
asymmetry is not responsible of the positive value of the social influence index $I_{SI}>0$ for 
the entire set of data. On one hand both subsets has about the same value of the uniform probability 
(i.e. $Y_u=5.72 X$ for the under-median set and $Y_u=6.26 X$ for the over-median set), and on the 
other hand, both subsets behaves similarly in the range of $X$ where both subsets overlap, even when 
the moving average for the  over-median subset is slightly shifted downwards suggesting that 
the threshold to hang a flag in average increases slightly with the household income.

We have also checked the correlation between flag density and the fraction of electors voting for each of 
the pro-referendum parties. As stated above, the Pearson's correlation coefficient for the 
data points in Figure (\ref{fig:BoV-VS-SoE}) is $r=0.64$, whereas for the data points plotted in planes 
$\nf/\nh - n_{SV,x}/\nre$ with $x$:  CIU, ERC, ICV, CUP and SI the values of r are 0.38, 0.68, 
0.26, 0.58 and 0.42 respectively. Figure ((\ref{fig:cincopartidos}) shows how the data points distribute 
in the these planes. For reference the upper-left panel is a repetition of Figure 
(\ref{fig:BoV-VS-SoE}). Note that we don't know to which party a household vote and therefore we 
can't know by simple observation how many of the flags posted in a ED can be attributed to each party 
sympathizing. Thus, the $Y$ axis in all panels includes the flag posted by households voter to all 
pro-referendum parties and therefore it is expected a degradation of the correlation between Y and 
the variable $n_{SV,x}/\nre$ for any of the $x$ pro-referendum parties. This is true for all cases 
except for the ERC case ($r=0.68$) which has a correlation coefficient slightly larger than the one 
in the consolidated case ($r=0.64$). A possible explanation to this pattern is that ERC voters tend 
to be the first to hang flags because they have the lowest thresholds and therefore act as 
influencer to other householders having higher thresholds, and consequently producing the higher 
correlation for $Y(X_{ERC})$. However, it is important to remark that the data at hand is not enough 
to validate this hypothesis. Regarding the positive curvature of the moving average note it is more 
evident when the $X$-axis includes the votes to all the pro-referendum parties (upper-left panel) 
than when it includes only the votes to one of these parties. Again, only the case for ERC shows 
a similar, but less pronounced positive curvature as the consolidated case shown in the upper-left 
panel. This reinforces the idea that when deciding to display a flag what matters is the number 
of displayed flags and/or the fraction of pro-referendum voters regardless of their sympathy for a 
particular pro-independence party.

\begin{figure}                            
\hbox{\includegraphics[scale=0.7,angle=0]{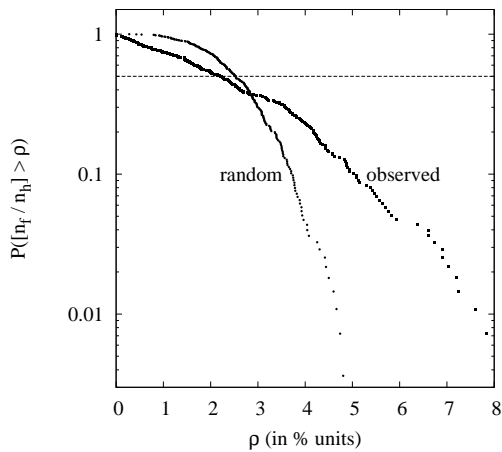}}
\caption{\label{fig:Pgtx}{\small
{\bf Probability $P(\frac{\nf}{\nh}>\rho)$ that the percentage of flags in an ED is higher than $\rho$}. As indicated by 
the labels, the two curves correspond to the probability distributions for the observed sample of EDs and for a 
simulation where flags are placed with the uniform probability $p_u=0.06075$ (see text). 
The horizontal line at $Y=0.5$ intersects the curves at their median values, 2.20\% for the observed sample 
and 2.56\% for the random sample.
}}
\end{figure}

Finally, Figure (\ref{fig:Pgtx}) compares the distribution $P(\nf/\nh > \rho)$ that gives the fraction of EDs having 
a covering percentage of flags higher than a given value $\rho$ for the observed distribution and for a randomly simulated 
distribution. The simulated distribution is created by randomly placing flags with the uniform probability $p_u=(\Nf/\Nh)/(\Nsv/\Nre)=0.06075$
in each of the $\nh(i) \times \frac{\nsv(i)}{\nre(i)}$ balconies in each of the 276 EDs. It is clear that the 
distribution $P(>\rho)$ of the actual data deviates significantly from a random one. There is an excess of EDs with 
both too few and too many flags. This deviation is in agreement with the finding presented above that low (high) 
density of flags inhibits (stimulates) the placement of new flags, which is consistent with a process of social influence.

In this first study the unit of observation was the ED and we demonstrate that the flag coverage observed in 276 EDs 
can not be explained by a linear correlation with the fraction of pro-referendum householders. On one hand, the moving 
average shown in Figure (\ref{fig:BoV-VS-SoE}) has a mean slope that clearly exceeds the one expected 
for a linear correlation with the fraction of pro-referendum householders ($I_{SI}>0$). On the other hand, the dispersion 
of coverages in EDs with similar fraction of pro-referendum householders can not be explained by the statistical 
fluctuations expected in a random process as demonstrated in figure (\ref{fig:Pgtx}). Moreover,   a 
randomly produced sample as the one in figure (\ref{fig:Pgtx}) can not explain the occurrence of 9  EDs with  zero 
flags as observed. All these facts strongly suggest that a mechanism of social influence is shaping the flag coverage 
patterns at the ED scale. 

\section{Flag dynamics at the microscale (household level)}{\label{sec:micro}}

\begin{figure}
\hbox{\includegraphics[scale=0.6,angle=0]{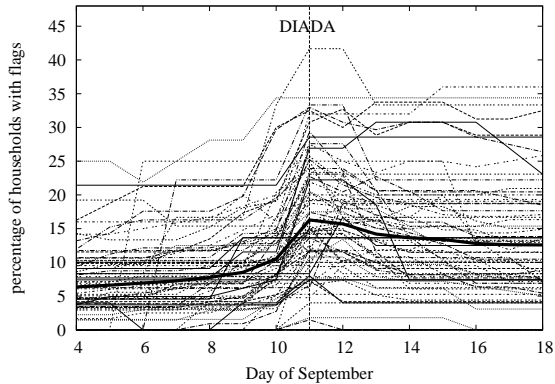}}
\caption{\label{fig:densities}{\small
{\bf Evolution of the density of flags in each of the 82 blocks of the sub-sample during the two weeks around 
the 2013 Diada}. The thick curve corresponds to the average density of flags in all the observed households. 
}}
\end{figure}

For our second study we observed the evolution of the detailed distribution of flags at balcony resolution 
in a sub-sample of 82 blocks' facades grouped in 16 different areas. The observation took 
place daily from September 4 to September 18 (i.e. two weeks around the national day of Catalonia, 
the Diada) followed by two additional observations, one on November 19 and the other on December 20. 
Among the 4,817 households observed, 918 showed a flag at least once. 
The evolution of the number and distribution of flags in the sub-sample is strongly case dependent. 
Figure (\ref{fig:densities}) shows the evolution of the percentage of observed households showing 
a flag for each of the 82 blocks. The black thick curve corresponds to the average density of flags 
in all the observed households. The Diada effect is reflected in a significant bump on the density 
of flags around September 11. However, there is a large dispersion in the density of flags among 
facades at any date and the pattern of evolution in each case is very diverse producing a large number 
of curve intersections. This case dependent behavior is indicative of the complexity of the phenomena under study.
Several factors may be involved in this dynamics. First of all, the initial distribution of flags 
on each block is different. Additionally, it is likely that the potential flaggers differ each other 
in how much social pressure is enough for them to hang (or remove) a flag, and therefore, due to 
the finite size of facades, large differences in the composition of flaggers in different blocks are 
to be expected. Also, the finite size of facades (and similar physical factors such as street width, 
building height, or balconies' visibility) may produce large differences on the particular 
distribution of flags on the facades. However, in any case we expect that short range mechanisms 
of social influence are affecting that distribution. But it is likely that the heterogeneities 
described above together with the global stimulus associated to the Diada event tend to blur the 
effect of local imitation. In the following we first characterize the time evolution of the total 
number of flags in the sub-sample and then we find evidences of very local social influence  by 
quantifying the departures of the clustering of flags in the observed data from an equivalent random 
distribution of flags. 

\subsection{Time Evolution}
\begin{figure}[ht]
\hbox{\includegraphics[scale=0.8,angle=0]{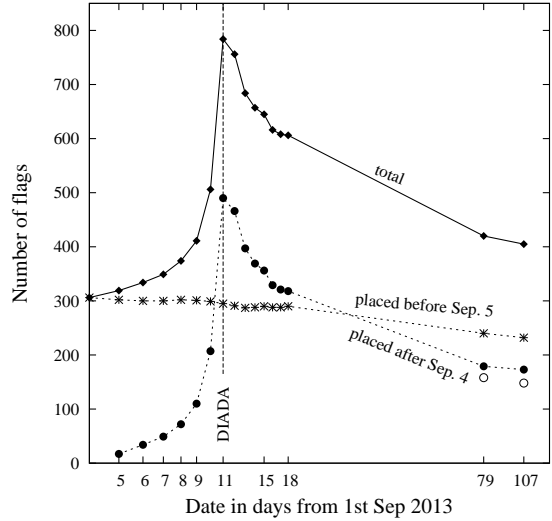}}
\caption{\label{fig:flgs_2sets}{\small
{\bf Time evolution of the number of flags exposed in the 82 blocks of the sub-sample}. The horizontal 
axis is logarithmic and indicates the date of observation measured in days from 1st September 2013. 
Days 79 and 107 correspond respectively to the observations made on November 19 and on December 20. 
As indicated in the labels, the square symbols represent the total number of flags, the starred 
symbols represent the number of flags in the households that had a flag the first day of observation, 
and the filled circles represent the flags hanged any day after September 4. The two open circle  
below the filled circle at the right end (t = 79 and 107) give the number of flags exposed on 
November 19 and on December 20 excluding the 21 flags that first appeared in the November 19 observation 
and the eight flags that first appeared in December 20.
}}
\end{figure}

Figure (\ref{fig:flgs_2sets}) shows the evolution of all flags exposed in the facades of the sub-sample 
during the 17 observation days. The upper curve ("total") shows the variation in time of the 
flags exposed in all the observed households, while the two lower curves show the evolution of 
the number of flags in those households where a flag was hanged for the first time before September 5 and 
after September 4, respectively. The first one, steadily decreasing, shows the evolution of the number 
of flags in the 306 households that had a flag the first day of observation (September 4). The 
second, up-and-down curve shows the evolution of the number of flags in the 612 households that 
hanged a flag for the first time from September 5 to December 20.  During the whole observation period, 
20 of the 82 blocks displayed zero flags at least one day, among which 9 never displayed any flag. 
Although the separation of the sample into these two subsets is arbitrary, it is guided by the fact 
that in the previous dates to the Diada there was a strong general stimulus to support the Catalan 
secessionist process by exposing a flag.  Within our observation period, the Diada effect seemed 
to induce 582 households to hang a flag.  Let us call the first subset of households `long term 
flaggers' (LTF), since they tend to keep their flag hanged for a long period of time. Note that 
during the 103 days separating the first and the last observation, the flags in this group decreased 100 
(232-306) / 306 = 24 \% while for the second group the number of flags decreased 
100 (582-173) / 582 = 70\%, so they may be called `short term flaggers' (STF), even when some (~30\%) 
of the people in the second subset may have become now LTF.
The steady decreasing curve for LTF can be roughly fitted by an exponential decaying function with 
a characteristic time scale of about one year. That is,
\begin{equation}
  b_{\leq4}(t) \simeq 306\, e^{-(t-4)/365} 
 \end{equation}
where t is the time in days counted from September 1. The bump due to the Diada event can be fitted as
 \begin{equation}
b_{>4}(t) \simeq  \left\{ \begin{array}{rll}
   17 \, & e^{+(t-5)/1.79}  & ~~~~{\rm if }\,\,   5\leq t \leq 11\\
   490 \, & e^{-(t-11)/14} & ~~~~{\rm if }\,\, 11\leq t \leq 25\\
   180 \, & e^{-(t-25)/365} & ~~~~{\rm if }\,\,  t > 25\\
\end{array}
\right.
\label{eq:x}
\end{equation}
where the first segment (September 5 to 11) corresponds to a rapid exponential growth with a characteristic 
growing time of about 2 days. The second segment (September 11 to 25) corresponds to a 
moderate decline with a characteristic decay time of about two weeks. The third segment represents a slow 
decrease after September 25 with again a characteristic declining time of about one year. The 
fit to this last segment is very uncertain due to the scarcity of observations in this period of time, 
but if true it means that among the 490 households that hanged a flag during the previous week to 
the Diada, about 180 became LTF.  
The evolution of the number of flags provides relevant information on the heterogeneity of the conditions 
under which people residing in the sub-sample decide to show their support for the process. 
However, it should be noted that some of the 63 flags that were hanged for the first time on September 11 
(the Diada) and were removed the next day are not necessarily supporting the secessionist 
process, since Catalans traditionally hang a flag that day long before the process started. 
Likewise, the 48 flags that were removed from the balconies the same day and reappeared the next day most 
likely belonged to people that took these flags to participate in the public demonstrations of the Diada.

\subsection{Departures from an equivalent random distribution}
The results obtained for the distribution of flags at the electoral districts level shown that the 
decision to hang a flag is not independent of others' decisions but subject to 
social influence. At the micro-scale, social influence must also be reflected in the spatial 
distribution of flags on the buildings' facades. One expects the effect of social influence to be 
inversely proportional to the separation between the flag and the potential imitator. If this is 
the case, flags should have a tendency to appear together in clusters. 

\begin{figure}
\hbox{\includegraphics[scale=0.3,angle=0]{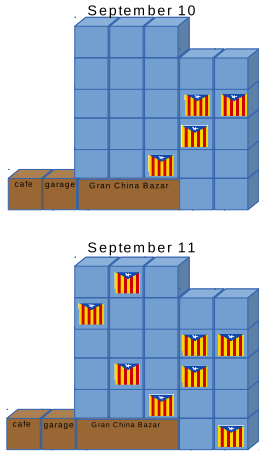}}
\caption{\label{fig:k59d9y10} {\small
{\bf Distribution of flags  on one of the smallest facades in the sub-sample in two consecutive days}. 
We assume that the distance between horizontally or vertically adjacent balconies is one unit and 
therefore, as indicated in Table 2, the average of the minimal distances between flags the 10-th day is 
$1.104= (1+1+1+\sqrt2)/4$ and the 11-th day is $1.362=(1+1+1+4\sqrt2+\sqrt5)/8$.
}}
\end{figure}

There are many ways to characterize the spatial distribution of a set of points. In order to detect clustering 
we use the average of the minimal distances between the points $\lambda$ (see figure (\ref{fig:k59d9y10}) and Table 2) as a 
simple way to estimate the tendency of flags to be close to one another; we have also tested using pair correlation functions, 
but since facades differ very much in shape and size it is convenient to use a scalar measure as $\lambda$. 
Let $\lambda_{\rm obs}(k,t)$ be the average minimal
distance between the $n(k,t)$ flags hanged in block $k$ (where $1 \leq k \leq 82$) during 
the date t (with $t=$September $4, 5, ... 18$, November 19 or December 20). However, $\lambda_{\rm obs}(k,t)$ can only 
be defined in facades with at least two flags (i.e. cases where $n(k,t)\geq2$).

 \begin{table}
\begin{center}
\begin{tabular}{c}
\hspace{0.0cm}{\bf Table 2} \\ Clustering index for the facades in Figure \ref{fig:k59d9y10}\\
\end{tabular}\\
\begin{tabular}{ccccccc}
\hline
\hline
 Day  &\vline &flags/households&$\lambda_{\rm obs}$&$\lambda_{\rm ran}$&$\sigma_{\rm ran}$&$C$\\
&\vline&---------------------&---------&---------&---------&---------\\
September 11& \vline  & 4/25 &  1.104 &  1.723 &  0.442 &  +1.446  \\
September 12& \vline  & 8/25 &  1.362 &  1.244 &  0.151 &  -0.778 \\
 \hline
\end{tabular}
 \end{center}
\end{table}

\begin{figure}
\hbox{\includegraphics[scale=0.6,angle=0]{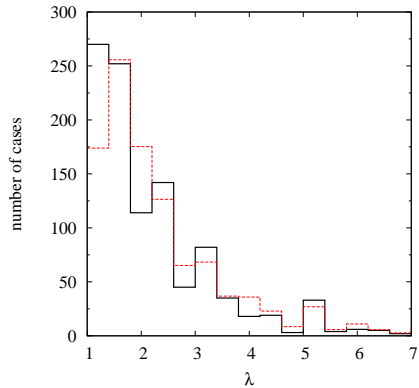}}
\caption{\label{fig:dmin-all} {\small
{\bf The continuous line histogram shows the distribution of the average minimum distance $\lambda$ for 
all the 1,040 observations (blocks) where there were at least 2 flags}. 
As a reference the red line histogram shows the distribution of $\lambda$ for $1040 \times 10000$ random simulations (see text). 
The width of each bar is 0.4. Note that 
distances between flags can be 1, $\sqrt2$, 2, $\sqrt5$, ... and the value of $\lambda$ 
is the average of the minimal distances between the $n(k,t)$ flags on a given block.
}}
\end{figure}

Note that we are looking for a clustering tendency in a very heterogeneous set of facades. Through all 
the observations (1,394: 82 blocks observed 17 times) there were 231 cases of blocks with no
flags hanged and 123 cases of blocks with only one flag hanged. In fact, there are 14 blocks 
where during the entire period of observation there were less than two flags. In total, $\lambda_{\rm obs}(k,t)$ 
can be calculated in 1,040 cases. The continuous line histogram in Figure (\ref{fig:dmin-all}) 
corresponds to the distribution of these 1,040 values of $\lambda_{\rm obs}(k,t)$. For comparison, the red 
histogram in Figure (\ref{fig:dmin-all}) shows the expected distribution for a random distribution of 
flags for these 1,040 cases. That is, for each block k and date t we randomly distribute $n(k,t)$ flags in the 
balconies/windows of an hypothetical facade with the shape of block $k$. We repeat this 
procedure 10,000 times for each block $k$ and date $t$ and calculate the average of the minimal distances 
$\lambda(k,t,i)$ for each of the random distributions $1\leq i \leq 10000$. The red histogram 
represents the distribution of these 1,040 x 10,000 values of $\lambda(k,t,i)$; the histogram is scaled to 
be appropriately compared to the distribution of the 1,040 observed values of $\lambda_{\rm obs}(k,t)$. 
Note that the number of observed cases with $\lambda$ values in the first bar is about 1.56 times 
the number expected for the random distribution. This suggests that some mechanism is enhancing the 
probability of occurrence of adjacent flags; however, the effect can be blurred because facades 
that differ very much in shape, size and number of flags can fall in the same bar. 
For example, a large facade with only 2 adjacent flags or a small facade full of flags both fall in the first bar, 
but the probability of having $\lambda=1$ in the former case is very small whereas in the later case the it is 1.

In order to circumvent this problem we homogenize the information introducing a `flag clustering index' defined as
\begin{equation}
C_{\rm obs}(k,t)=\frac{\lambda_{\rm ran}(k,t)-\lambda_{\rm obs}(k,t)}{\sigma_{\rm ran}(k,t)}
\label{eq:Cobs}
\end{equation}
where the quantity 
\begin{equation}
\lambda_{\rm ran}(k,t)=\frac{1}{10000}\sum_{i=1}^{10000} \lambda(k,t,i)
\label{eq:lrankt}
\end{equation}
is the mean value of the average minimal distances 
for 10,000 random distributions and $\sigma_{\rm ran}(k,t)$ is its standard deviation. Note that with a different meaning, 
the term clustering index is used for the characterization of networks and data structures. 

For a given block $k$ and date $t$, the clustering index $C_{\rm obs}(k,t)$ measures in standard deviation 
units the departure of the observed distribution of the $n(k,t)$ flags from the average distribution of a 
large number of random sets. Note that as the density of flags on a facade increases, the average minimal distance tends 
to decrease, but also the standard deviation decreases. Figure (\ref{fig:k59d9y10}) shows one of the smallest facades in 
the sample in two consecutive days. The shape of the facade has been altered to prevent identification of the households.
day before the Diada there were 4 flags and the next day 4 more appeared. Table 2 shows the 
corresponding values of $\lambda_{\rm obs}$, $\lambda_{\rm ran}$, $\sigma_{\rm ran}$ and $C$ for these two days. 
This is just an example that is not representative at all; there are blocks that are made 
of a single large rectangular building, whereas others have several buildings of different size. 

\begin{figure}
\hbox{\includegraphics[scale=0.7,angle=0]{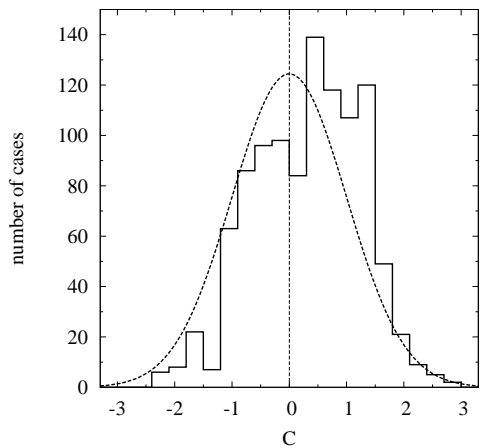}}
\caption{\label{fig:distDeltamintodos}{\small
{\bf The histogram represents the clustering index $C_{\rm obs}$ as defined in equation (\ref{eq:Cobs}) for the 1,040 observed facades}. 
 The width of bars is 0.3. The dotted curve is the normal distribution ($0.3 \times 1,040 \times \frac{e^{-x^2/2}}{\sqrt{2\pi}}$)  
 expected  for an equivalent random distribution of flags. The fact that $C_{\rm obs}$ shows a marked excess of cases with $C_{\rm obs}>0$
 is a clear indication that globally the hangings of flags are not independent events but that a local influence 
 mechanism is favoring their clustering.
}}
\end{figure}

Figure (\ref{fig:distDeltamintodos}) shows the distribution of the 1,040 values of $C_{\rm obs}(k,t)$ and 
the expected normal distribution for an equivalent random placement of flags as described 
before \footnote{We have verified that a normal distribution centered at zero with $\sigma=1$ is an excellent representation 
of the distribution of the clustering indexes of a large set of random placements of $n(k,t)$ flags on a facade with the shape 
of the block $k$. The clustering index for a particular random placement $i$ of $n(k,t)$ flags is 
$C_{\rm ran}(k,t,i)=\frac{\lambda_{\rm ran}(k,t)-\lambda_{\rm ran}(k,t,i)}{\sigma_{\rm ran}(k,t)}$.  
Note that by construction the mean value of $C_{\rm ran}(k,t,i)$ for each block $k$ and each date $t$ is zero 
and therefore the mean value of $C_{\rm ran}$ for all the 1,040 cases is also zero.}.
Inspection of Figure (\ref{fig:distDeltamintodos}) shows that the distribution of the 1,040 values of $C_{\rm obs}(k,t)$ (black histogram) 
is clearly shifted toward positive values, indicating that in overall the flags on the observed facades are markedly more clustered than 
expected for an average random distribution. The percentage of cases with $C>0$ is 63\%.

\begin{figure}
\hbox{\includegraphics[scale=1.0,angle=0]{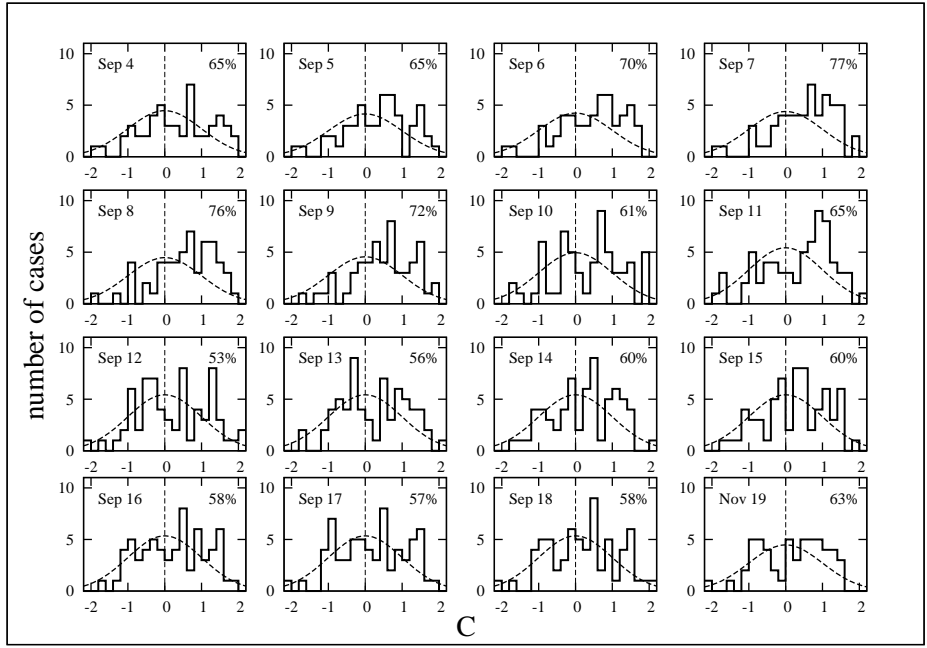}}
\caption{\label{fig:distDeltaDay} {\small
{\bf Same as Figure (\ref{fig:distDeltamintodos}) with data disaggregated by day of observation.} Bin width is 0.2. 
The labels give the percentage of cases with $C_{\rm obs}>0$ and the day of the observation.
}}
\end{figure}

The same analysis is repeated but for the observations in each single day.  
Figure (\ref{fig:distDeltaDay}) shows the frequency of occurrence of the daily values of $C_{\rm obs}$ 
in the blocks with at least two flags. The 
labels in each plot indicate the date of observation and the percentage of blocks with $C_{\rm obs}>0$. 
The results for the observations made on the 20 December are not included in figure 
(\ref{fig:distDeltaDay}) because they are very similar to the results for the 19 November shown in the bottom-right plot. 
Again, the black histograms correspond to the distribution of $C_{\rm obs}$ 
and the dotted curve is the normal distribution expected  for an equivalent random distribution of 
flags for that day. All plots in Figure (\ref{fig:distDeltaDay}) 
show again a clear excess of flag distributions with $C_{\rm obs}>0$.

\begin{figure}
\hbox{\includegraphics[scale=0.65,angle=0]{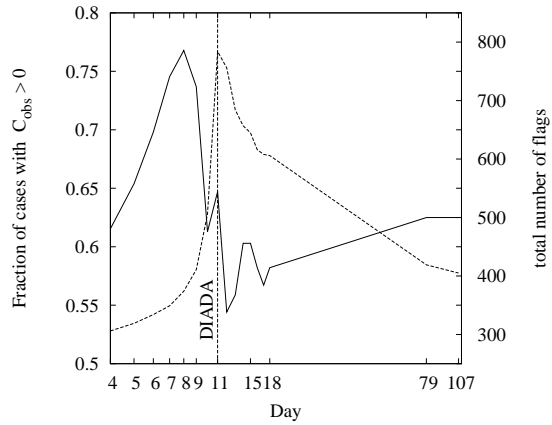}}
\caption{\label{fig:fracover0} {\small
{\bf Evolution of the overall fraction of cases with $C_{\rm obs}>0$ (continuous curve; left- hand Y axis) and the 
total number of flags (dashed curve; right-hand Y axis)}.
}}
\end{figure}

Figure (\ref{fig:fracover0}) shows the evolution of the fraction of blocks with $C_{\rm obs}>0$. 
A strong fluctuation of the clustering is associated to a notorious increase of the number of flags 
due to the Diada effect. From 4 to 9 September the number of flags increases from about 300 to 400 and the 
clustering index increases about 10\%; in the next three days the clustering index decreases 
about 20\%, and during the following weeks it progressively recovers the pre-DIADA values.

\begin{figure}
\hbox{\includegraphics[scale=1.0,angle=0]{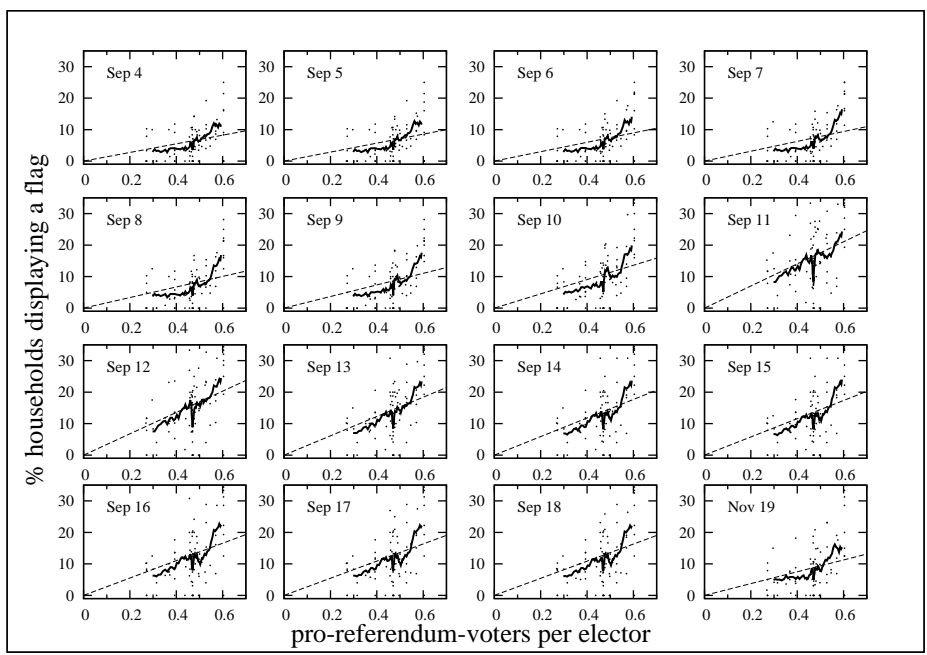}}
\caption{\label{fig:BoV-VS-SoE_dia}{\small
{\bf Same as figure (\ref{fig:BoV-VS-SoE}) but for the flags in the facades of the 82 blocks of the  second study}. 
The labels indicate the day of observation. The fraction of pro-referendum voters is 
assumed to be the same as the ED to which the  block belongs
}}
\end{figure}

Finally we have verified that the facades observed in the second study shows a similar pattern in the 
plane $n_f/n_h$ VS $n_{vs}/n_{ele}$ than the electoral districts observed in the first study. 
Figure (\ref{fig:BoV-VS-SoE_dia}) shows the  percentage of flag per household measured in the 82 blocks as a function 
of the fraction of  pro-referendum voters. The information about the electoral 
results is available for EDs but not for facades, however we assume that the facades has the same electoral 
behavior of the EDs to which they belongs since large differences in electoral behavior 
among the facades in a ED are not expected. Even when the  size of the sample in the second study is much 
smaller (82 cases, with 60 households per block in average) compared with the first study (276 
EDs, with about 770 households per ED), it is very interesting that the observed patterns in both 
cases follows the same trends. The first study was performed July 2013 during a steady political 
period in the sense that there was no special event that impacted the number of flags in the city. 
Therefore, the most suitable dates of the second study to compare the with the July pattern are 4 
September, 19 November and 20 December. 
Comparison of figures (\ref{fig:BoV-VS-SoE_dia}) and (\ref{fig:BoV-VS-SoE}) shows that in most panels the 
moving average curves has a similar shape and crosses the uniform probability lines in a similar way than 
in the first study. In fact, only the dates two or three days around the Diada display patters 
with low indexes $I_{SI}$. It is worth mentioning that the uniform probability $p_u$ in both studies can not
be directly compared since on one hand the number of flags in the city was not the same, and on 
the other hand the number oh households in the first study includes residences without balcony-window in 
the external facades of the ED, whereas in the second study we counted directly the number of 
households in each facade.

The results of this second study shows that the effect of social influence is affecting both, 
the distribution of flags in a facade (excess of blocks with positive clustering index) and the pattens of 
the distribution of blocks in the plane $n_f/n_h$ VS $n_{vs}/n_{ele}$. It is probable that the 
same micro-mechanism is responsible of these two effects even when the first manifest at the 
micro-scale whereas the second manifest at the meso-scale. Additionally, the fact that the observed 
number of cases with zero flags (9 EDs in the first study and 20 facades in the second study)
largely exceeds the number expected by random indicates that there is a strong inhibitory social 
pressure against being the first to publicly manifest your political preference in your environment.
In a forthcoming paper we consider an agent-based model to test, among others, these hypothesis.

\section{Discussion and conclusions}{\label{sec:conclusions}}
We have documented and characterized a process of social influence in the expression of political 
preferences departing from direct observation in the field. Two observational studies on how people 
hang pro-referendum flags in Barcelona in the context of the Catalan secessionist process provide 
data to support the claim that the probability that a private household's balcony or window shows a 
flag when the inhabitants have secessionist preferences is significantly affected by how many neighbors 
at the local level  have hung  a flag. The first study at the ED level shows that there is an 
inhibition-stimulation dynamics in flag-hanging, depending on whether pro-referendum voters are or not 
majority in each ED. Departing from the correlation between flag density and vote for pro-referendum parties, 
flags appear more frequently than expected in districts with a majority of pro-referendums 
voters, and less frequently than expected in districts where such a majority does not exist (in fact, 
the excess of ED with zero flags is also indicating that a social inhibition effect is operating 
in the latter districts). Since the social influence mechanism at ED scale is likely to be visually mediated, 
a more precise reasoning would be that pro-referendum voters has an enhanced tendency to 
hang a flag when they observe that in their environment there is a high density of flags, which usually, 
but not necessarily, occurs in EDs with high proportion of pro-referendum voters. This is in 
line with Latan\'e's \cite{Latane81}
dynamic social impact theory, which predicts that an individual will be more 
likely to conform to the preferences and behavioral propensities of the local numerical majority, 
and that this produces the clustering of attitudes and behaviors. The second study at the balcony-window 
level allows to isolate short term flaggers from long time flaggers and to show that social 
influence is also reflected in how flags are distributed on the buildings' facades. We calculated a clustering 
index in order to show that flags tend to appear more clustered than expected by chance. 
The clustering index  adopts positive values when the average minimal distance between flags in a facade 
is smaller than the value expected by random, and we find that there is a clear excess of 
facades having positive values implying that the effect of social influence includes a very local term that 
reaches its maximum when the potential flagger sees a flag flying at his side. The excess of 
positive clustering index values  persist for all the dates observed showing that the successive hangings of 
flags are not independent events but that a local influence mechanism is favoring their 
clustering. (again a prediction of Latane's dynamic social impact theory). It is somewhat surprising that 
the effect of local social influence manifests so clearly in this second study as well as in 
the first one (see Figure (\ref{fig:BoV-VS-SoE})), when certainly other non-local mechanisms of social 
influence may play a role in the decision of displaying a flag, such as mass media information, 
political discussion with acquaintances living in other EDs, or the observation of the flag density in 
other areas of the city. It is out of the scope of this work to quantify the relative importance 
of local, non local and global processes of social influence, but the results described above indicate 
that local social influence is important in determining the social attitudes that individuals 
adopt in such an environment.
Alternative explanations usually confounded with social influence \cite{Manski00}
such as homophily could hardly explain the observed patterns. At ED scale a possible effect
of homophily in terms of income level is discarded since the inhibition-stimulation dynamics 
persist when controlled by this variable (see Figure \ref{fig:BoV-VS-SoE_2rentas}). At
block scale a possible effect of clustering produced by the segregation of the potential flaggers
in terms of their thresholds to hang a flag is highly unlikely. On one hand, people residence's decisions 
were taken generally long time ago, and on the other hand, the information about neighbors' tendency to 
publicly express their political preferences is not a relevant factor to take the desicion and 
moreover is not accessible.
The possibility of differentiated contextual effects can also be ruled out since the Catalan process 
and the related political struggles and discussions to which all residents in the city are exposed 
are essentially the same.
A limitation of the studies presented is that they do not allow inferring the specific type of 
social influence mechanism operating. We show that some kind of social influence is necessary to explain 
the data, but we do not identify concrete mechanisms or dynamics. This is a usual shortcoming in social 
influence studies, since different mechanisms are typically compatible with observed patterns of 
influence \cite{LopezEA08}.
However,  the two studies presented here provide detailed quantitative information 
that can be used to constraint the functional form of the mechanisms of 
social influence that drive the flag dynamics. The more suitable approach to test different 
mechanisms is through agent-based simulations since these numerical models allows to produce simulated 
patterns that can be compared to the observed data through characterizing quantities such as 
e clustering index C, the social influence Index $I_{SI}$ and the correlation coefficient r, as well as 
the local and global evolution of the number of flags. In a forthcoming paper we will 
present the results from an agent-based model that includes various mechanism of social influence acting at 
various scales. However, it is not necessary to perform a numerical simulation to envisage that 
to reproduce a systematic excess of blocks with positive clustering index is necessary to include a 
very local process of social influence that enhance   pro-referendum-voters to hung a flag near 
the existing ones; it would be hardly justifiable that some homophilic process has produced a 
segregation of pro-referendum-voters in the block facades that in turn induce the observed 
clustering of flags. Since data on psychological or attitudinal dispositions to hang flags were not 
available, the direct observation of flag-hanging behavior is not enough to identify specific types of 
social influence. However, although there may be different psychological mechanisms operating, 
they all lead to a similar behavioral rule: do A if enough members of the relevant group do so. 
We have provided evidence that this behavioral rule is operating in our case.

\section*{Acknowledgments}
this work benefited from financial support by the National Plan for R+D+I of 
the Spanish Ministry of Science and Innovation and the Spanish Ministry of Economy and Competitiveness 
through Grants CSO2012-31401 and CSD 2010-00034 (CONSOLIDER-INGENIO). Paula Hermida benefited from a 
FPU pre-doctoral fellowship granted by the Spanish Ministry of Education. Arnau Bàguena, Adri\'an 
Becerra, Xavier Bosch, Lluc Cah\'is, Indira Centellas, Marc Collado, Joan Gasull, Alex Gim\`enez, Xavier 
Guijarro, Alejandro L\'opez, J\'ulia de Quintana and Irene Rodr\'iguez participated in the fieldwork. We 
are especially grateful to J\'ulia de Quintana and Álex Gim\'enez for their valuable support in technical 
tasks and in coding the data for the second study. We thank the Department of Statistical 
Information of Barcelona's City Hall for their help in obtaining some of the data for the study. 
Finally, we want to thank Javier G. Polavieja and other attendants to the VII Conference of the 
International Network of Analytical Sociology (INAS, Mannheim, June 2014) and the V Econosociophysics 
Workshop (IPHES-URV, Tarragona, January 2014) for their valuable comments.

\end{document}